\documentclass[review]{elsarticle}

\usepackage{lineno,hyperref}
\modulolinenumbers[5]

\journal{Journal of \LaTeX\ Templates}

\begin{document}

\begin{frontmatter}

\title{How the inter-electronic potential Ans\"{a}tze affect the bound state solutions of a planar two-electron quantum dot model}


\author[mymainaddress,mysecondaryaddress]{F. Caruso\corref{mycorrespondingauthor}}
\cortext[mycorrespondingauthor]{Corresponding author}
\ead{caruso@cbpf.br}

\author[mysecondaryaddress]{V. Oguri}
\author[mysecondaryaddress]{F. Silveira}

\address[mymainaddress]{Centro Brasileiro de Pesquisas F\'{\i}sicas -- Rua Dr.~Xavier Sigaud, 150, 22290-180, Urca, Rio de Janeiro, RJ, Brazil}
\address[mysecondaryaddress]{Instituto de F\'{\i}sica Armando Dias Tavares, Universidade do Estado do Rio de Janeiro -- Rua S\~ao Francisco Xavier, 524, 20550-900, Maracan\~a, Rio de Janeiro, RJ, Brazil}

\begin{abstract}
The model of a two-electron quantum dot, confined to move in a two dimensional flat space, in the presence of an external harmonic oscillator potential, is revisited for a specific purpose. Indeed, eigenvalues and eigenstates of the bound state solutions are obtained for any oscillation frequency considering both the $1/r$ and $\ln r$ Ans\"{a}tze for inter-electronic Coulombic-like potentials in 2$D$. Then, it is pointed out that the significative difference between measurable quantities predicted from these two potentials can shed some light on the problem of space dimensionality as well as on the physical nature of the potential itself.
\end{abstract}


\end{frontmatter}


\section{On the two Ans\"{a}tze}

How Physics is affected by space-time dimensionality is a general question that is gaining prominence due to strong developments in Unified Theories and in Planar Physics.

All the attempts trying to fix or understand space dimensionality~\cite{Petkov} which depends on the form of physical potential has to face an insurmountable epistemological problem, so far as spaces with higher dimensions ($D$) are considered, as suggested by many unified theories: that we are not able to probe such extra dimensions and, therefore, we cannot infer the generalized mathematical formula of the potential from experience in this $D$-fold space~\cite{Caruso2}.

To study a particular physical system in $D$-dimensional spaces with $D>3$ one should postulate the validity of the same equation that describes it in 3$D$. The only ``justification'' for this falls back on the expectation that some version of the Anthropic principle should be still valid in higher dimensions. This means that if life depends on a particular physical law in 3$D$, life should still exist in higher dimensions with similar dependence on this law~\cite{Life-Dimensionality}.

Thus, for $D>3$, the specific case of the Coulombic potential depends on two Ans\"{a}tze. The first (and more general) one, inspired in an early proposal from Ehrenfest~\cite{Ehrenfest1,Ehrenfest2}, is to assume the validity of Poisson's equation for the gravitational potential (in a classical level) in any higher dimension associated to the idea that such validity should be requested to assure planetary stability necessary to the maintenance of life as we know it. In the framework of General Relativity this problem was treated in~\cite{Tangherlini}, and the quantum hydrogen atom in higher dimensions was discussed in~\cite{Tangherlini,Hidrogen-atom}. Within this assumption, we still have to introduce another Ansatz concerning the mathematical form of the potential. There are two possibilities that will be called throughout this paper Ansatz 1 and 2. The first one admits that the $1/r$ behavior of the three-dimensional Coulombic potential is still the same, no matter the number of space dimensionality is considered~\cite{Nieto,Nieto2,Shaqqor}; and Ansatz 2, supposes that the potential depends on the dimensionality $D$ as $1/r^{D-2}$~\cite{Ehrenfest1,Ehrenfest2,Supplee,Mostepanenko,Jordan}, for $D\geq 3$, and as $\ln r$, for $D=2$. Ansatz 1 has the advantage of giving rise to some analytically solvable problems. However, Ansatz~2, in spite of an intrinsic mathematical difficulty, has the advantage of ensuring, at least at the classical level, the electric charge ($e$) conservation~\cite{Supplee}. Indeed, in $D$ dimensions, from the Poisson equation for the electric field, $\vec E = - \vec \nabla \varphi$, it follows the integral form of Gauss'  law

\begin{equation}
\int (\vec \nabla \cdot \vec E) \mbox{d}V = \int (-\nabla^2 \varphi) \mbox{d}V\ \sim e \quad \Rightarrow \quad \varphi \sim \frac{e}{r^{D-2}}
\end{equation}

Clearly, in the 2$D$ case, the validity of Poisson's equation need not to be postulated, removing the epistemological barrier mentioned at the beginning of this Section. In addition, for this space dimensionality, one can justify the use of the $1/r$ potential by arguing that the experimental set up is anyhow three-dimensional, even if some of its dimensions is very tiny. Those points are still open questions in Planar Physics and should be considered.

As a last remark, it is important to stress that calculations carried on this paper, for both Ans\"{a}tze, can help in given an answer to the question Ehrenfest puts to himself in 1920:
``What role does the three-dimensionality of space play in the fundamental laws of physics?''~\cite{Ehrenfest1}.
Following this general idea, Weyl thought about ``what inner peculiarities distinguish the case $n=3$ among all others?'', and then, considering the creation of the world, asked himself if  space was 3-dimensional from the beginning, and whether a `reasonable' explanation of this fact can be given by disclosing such peculiarities or not~\cite{Weyl}.
Indeed, precise measurement of the quantum dot energies can endorse the investigation of whether one really have to consider a $\ln r$-potential in 2$D$ or not. Besides that, such a comparison between theoretical prediction for different potentials and experimental values allows us to examine another important question: if there is a particular spatial scale associated to one particular dimension (a phenomenological limit) for which the quantum system can actually be considered as effectively a 2$D$-system although immersed in a 3$D$ space, or if only strictly two-dimensional systems have to be described by a $\ln r$-like potential.

This is a special and attractive possibility of discussing the existence of a phenomenological limit between two different physical descriptions of a system in 2$D$ or 3$D$. Notice that this is not a general rule. In fact, it is well known that the theoretical quantum description of a massless spin 1 particle could not follow from the $m\rightarrow 0$  limit of the theory for massive particles with the same spin quantum number. In any case, such hypothetical limit cannot even be experimentally tested in this example simply because there is no other massive spin 1 particle having rest mass between 770~MeV (for the $\rho$) and zero (for the photon $\gamma$). In addition, there is no physical process able to diminish the rest value of masses. Therefore, the fact that we can continuously reduce the length scale of one of the three dimensions of a particular system (as in the case of quantum dots) or film brings with itself new alternative possibilities.

\section{Planar Physics and the $ln(r)$ potential}\label{former_prediction}

Over the recent past years, planar Physics and truly two-dimensional systems have attracted a great deal of attention in connection with graphene mono- \cite{Geim1,Geim2} and bi-layers~\cite{Zhang}, and with Quantum Hall Effect~\cite{Laughlin,Prange}. All these quantum systems point to consider that the correct Coulomb inter-electronic potential should be the
$\ln r$-potential~\cite{Eveker}, rather than assuming the validity of the usual 3$D$-dependence in planar systems, corresponding to the $1/r$-potential~\cite{Priyanka,Sadeghi}.

In addition, if one investigates the statistical-me\-chan\-ics properties of a two-dimensional Coulomb gas, one must actually consider a logarithmic potential to get that the spectrum is purely discrete. The $1/r$-dependence is not compatible with this result. Now, that we are able to carry out highly precise experiments with genuinely planar systems, pursuing a deep investigation of the mathematical properties and physical features of the log-potential in 2$D$ is a relevant task. The scope of this paper is to shed light on these questions.

In particular, how the two-dimensional ln$r$-potential should affect the energy levels of the quantum dots~\cite{Reimann} deserves a careful investigation, which, to the best of our knowledge, was not yet carried out.

The 2$D$ radial Schr\"{o}dinger equation, in atomic units ($\hbar = m = e =1$), for the relative coordinate $r$, can be obtained by introducing the polar coordinates $(r,\theta)$ and putting its solution in the form
\begin{equation}
\label{main_sol}
\psi(\vec r) \propto r^{-1/2}\, u(r)\, e^{\pm i l \theta}
\end{equation}
where $l$ is the integer angular momentum quantum number of the two-dimensional system. The radial function $u(r)$ should satisfy the following equation:
\begin{equation}\label{radial_equation}
\frac{\mbox{d}^2u(r)}{\mbox{d}r^2} + \left[ \eta - V(r) - \omega^2 r^2 - \frac{(l^2-1/4)}{r^2} \right] u(r) \equiv \frac{\mbox{d}^2u(r)}{\mbox{d}r^2} + \left[ \eta - V_{\rm eff}(r) \right]
u(r) = 0
\end{equation}
where $\eta$ is the radial energy and the frequency $\omega = \Omega/2$.

The problem of two electrons in an external oscillator potential~\cite{Merkt} is studied in three dimensions~\cite{Taut_93}, and it is shown that the above radial equation is \textit{quasi-exactly solvable}~\cite{Turbiner_1, Turbiner_2, Usheridze}, which means that it is possible to find exact simple solutions for some, but not all, eigenfunctions, corresponding to a certain infinite set of discrete oscillator frequencies. In a recent paper~\cite{Caruso, corrigendum}, adopting the Ansatz $V(r) = 1/r$, it was shown that in 2$D$ it is possible to determine exactly and in a closed form a finite portion of the positive energy spectrum and the associated eigenfunctions for the Schr\"{o}dinger equation describing the relative motion of a two-electron system, by putting Eq.~(\ref{radial_equation}) into the form of a Biconfluent Heun Equations (BHE)~\cite{Ronveaux}.
The particular solutions $u_{nl}(r)$ can be obtained by analytically solving the BHE equation just for a discrete set of external frequencies $\Omega$~\cite{Caruso}. This method, indeed, did not give us a polynomial solution for any frequency $\Omega$; each solution is obtained for a specific frequency value. The same is true for other similar studies~\cite{Taut_93, Taut_94, Taut_2000, Taut_2010}.

Therefore, it is natural to wonder if a numerical analysis of this problem could give rise to a broad class of solutions well defined for any external frequency value $\Omega$ as it is expected in a real experimental situation. This take us to solve numerically the radial Schr\"{o}dinger equation, Eq.~(\ref{radial_equation}), by using the Numerov method~\cite{Numerov,Numerov2,Numerov3,Numerov4,Numerov5,Numerov6}.  The algorithm was implemented in a program developed by the authors using $C^{++}$ language. Both calculations and graphics shown in this paper were done by using the CERN/ROOT package. All these solutions, to the best of our knowledge, were unknown up to now. The success of this task opened new possibilities, as the calculation of the bound-state solutions for the two-electron quantum dots, which is done in this paper. The other possibilities, that will be treated elsewhere, are: the verification of all previous analytical results; the determination of new states for arbitrary $\Omega$ values; and, finally, the possibility of investigating the difference on both the eigenvalues and eigenfunctions by changing the Coulombic potential from $1/r$ to $\ln r$, as is expected for a strictly two-dimensional system.


\section{Bound state solutions for both Ans\"{a}tze}\label{3}

\begin{figure}[htb!]
\centerline{\includegraphics[width=130mm]{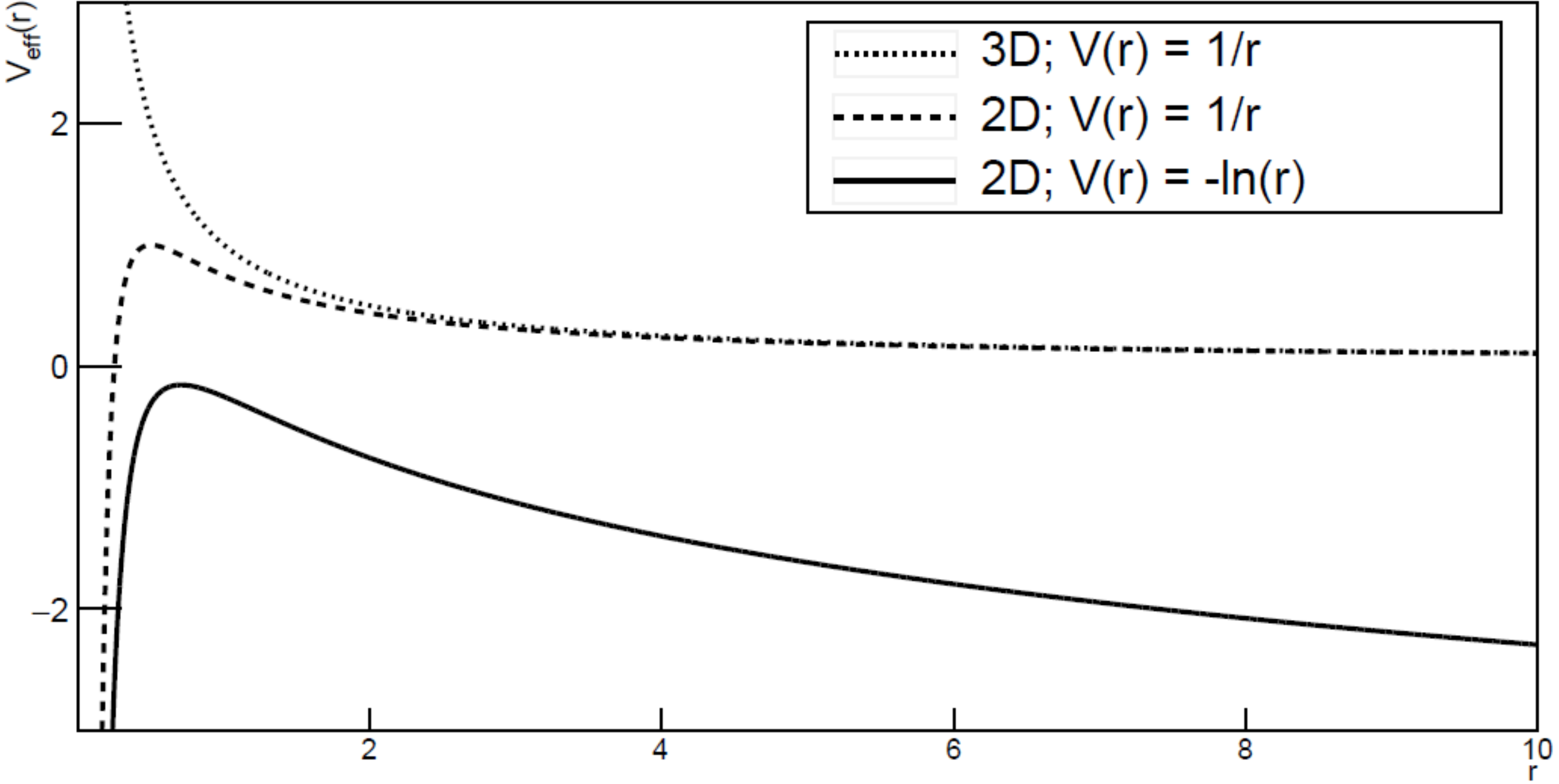}}
\caption{Effective potentials corresponding to different choices of space dimensionality and the Coulombic potential between two electrons.}
\label{fig.1}
\end{figure}

The very peculiar form of the two-dimensional geometrical (metrical) term $(l^2-1/4)/r^2$, in Eq.~(\ref{radial_equation}), allows the existence of a bound state solution corresponding to a $S$-wave ($l=0$), no matter the Ansatz for the potential is. This is a unique feature of Planar Physics, as can be straightforwardly seen from Fig.~\ref{fig.1}.

In 2$D$, for each kind of the effective potential, we were able to numerically find just one state solution. The ground state energy value computed from Eq.~(\ref{radial_equation}), is $E=-63.92$~Ha, for the $1/r$ potential, to be compared with $E=-45.92$~Ha, for the $\ln(r)$ potential. Both values are obtained with a typical value of the external frequency $\omega=0.01$~Ha. However, for small values of $r$ (between 0 and 0.5), the well shape is not at all bias by the $\omega$ choice (at least in the range we are considering in this paper). If these values are compared to the ground state energy of the hydrogen atom in three dimensions, $E_H = -0.5$~Ha, we see that they differ from two order of magnitude. Thus, the bound energy of the two-electrons in the quantum dot model considered here in 2$D$ is significantly greater than that of the hydrogen atom in 3$D$.

There is, in any case, a net difference between the two predictions which can be experimentally probed in a straightforward way.

\begin{figure}[htb!]
\centerline{\includegraphics[width=130mm]{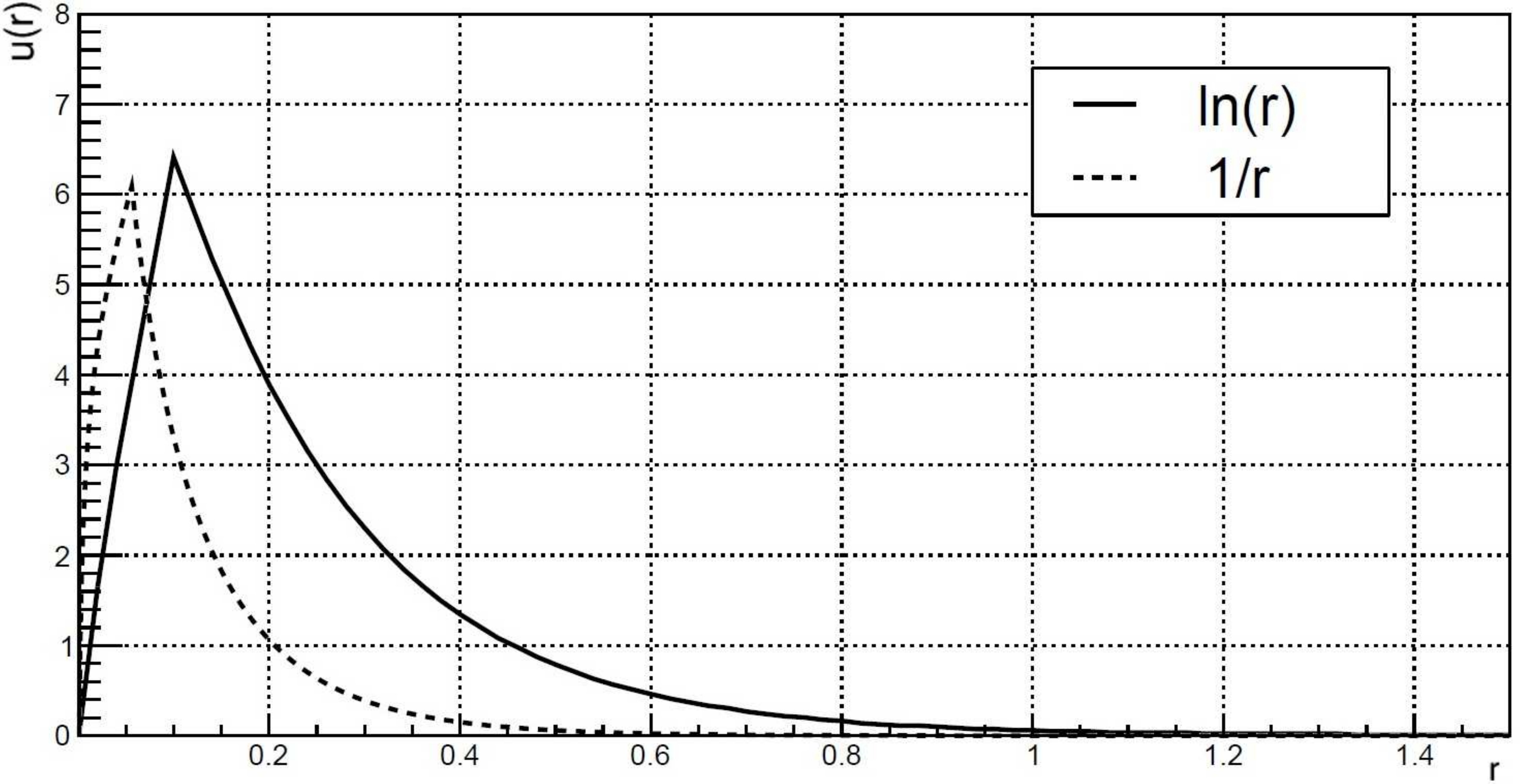}}
\caption{Comparison between wave-functions for the ground state in the case of a $\ln(r)$ and $1/r$ potentials.}
\label{fig.2}
\end{figure}

Just for completeness, the eigenfunctions corresponding to these two bound-states are given in Fig.~\ref{fig.2}.

This kind of discontinuity in the first derivative of the wave function seen in Fig.~\ref{fig.2} is typical of the well studied $\delta(r)$-potential. It should be stressed that in this case and even in that of a very deep well with a small characteristic width (which is indeed our case) only one energy eigenstate is expected in the framework of non-relativistic quantum mechanics, as we have found.

\section{Discussions}\label{discus}

We can still confront our result for the planar bound state energy solution for a quantum dot, $E=-45.92$~Ha, in the case of the $\ln(r)$ potential, with the two-dimensional bound state energy of a hydrogen atom assuming a Chern-Simons interaction~\cite{Jordan}, described by a potential energy
$$ V(r) =\frac{e^2}{2\pi} K_0\left(\frac{m_\gamma c r}{\hbar}\right)$$
where $m_\gamma$ is the photon effective topological mass. It should be stress that these two calculations follow the same methodology. Although in the limit of small values of $r$ the Bessel function behaves like $\ln r$, the value of the bound state energy in impressively different from what is found in the present paper. Indeed, the result found in~\cite{Jordan}~for the atom bound state energy is $E=-0.00013$~Ha, considering $m_\gamma/m_e \simeq 10^{-5}$. If we compare this result for the 2$D$ hydrogen ground state with the three-dimensional analogous energy $E_n \sim 1/n^2$, we see that such prediction for the ground state energy of a planar atom corresponds to the energy of a 3$D$ atom with the principal quantum number $n=62$. In other words, the Chern-Simons interaction gives rise to planar hydrogen atoms which behaves like a Rydberg atom. This result is qualitatively and quantitatively different from our prediction for the ground state energy of a planar quantum dot which was found to be nearly 90 times greater than the analogous energy state of the three-dimensional hydrogen atom.

As already pointed out, the significative difference between the bound state energy values can be tested experimentally and so used to shed light on the problem of space dimensionality, as mentioned. Indeed, if space is strictly two-dimensional, the modulus of quantum dot energy computed for the $\ln r$-potential will be 72\% of the corresponding value got with the $1/r$ potential.

The search for such bound state is important since it is a unique prediction of strictly planar-physics. In any case, we stress again that they are quite two order of magnitudes greater in modulus than the bound energy of the hydrogen atom in 3$D$.

From the experimental point of view, the spectrum of quantum dot junctions prepared with molecular beam epitaxy has been determined by applying electrostatic potential bias that allows electrons to tunnel through the device \cite{hill}.

A more promise technique to measure the spectrum of planar quantum dots involves the use of the electrostatic force microscope with single-electron sensitivity (e-EFM)~\cite{cok, roy}. Indeed, this technique is able to probe the spectrum of discrete energy levels of an electron from a quantum dot, even when they are shared by several degenerate
states~\cite{cok}. By adjusting the resonance frequency of a cantilever which motion modulates the charge state of a quantum dot junction, one can determine the tunneling rate of electrons across the junction. Besides the charge state of the device the technique provides also the dynamics of tunneling single electrons,  enabling quantitative spectroscopy measurement of energy level structure.

All this emphasize the interest in the study of two-dimensional physical systems and can be seen as a justification to pursue the investigation on the planar hydrogen atom and on the planar quantum dot.

\section*{References}

\bibliography{mybibfile}

\end{document}